\documentclass[aps,prx,twocolumn,groupedaddress]{revtex4-2}

\newcommand\figref{Fig.~\ref}
\newcommand{\kxNorm}{\overline{k}_x}
\newcommand{\kyNorm}{\overline{k}_y}
\newcommand{\kyNormIn}{\overline{k}_{in,y}}

\newcommand{\kyNormOut}{\overline{k}_{out,y}}

\newcommand{\etaxNorm}{\overline{\eta}_x}
\newcommand{\etayNorm}{\overline{\eta}_y}
\newcommand{\muTensor}{\boldsymbol{\overline{\overline{\mu}}}}

\usepackage{amsmath}
\usepackage{amssymb}
\usepackage{bm}% bold math
\usepackage{relsize}
\usepackage{xcolor}
\usepackage{graphicx}% Include figure files
\usepackage[caption=false,font=footnotesize]{subfig}
\graphicspath{{_FINAL_FIGURES_PMM_paper}}
\usepackage{tabularx}
\usepackage{comment}

\begin{document}

%\preprint{}

\title{Perfectly Matched Metamaterials}


\author{Jorge~Ruiz-Garc\'ia}
\altaffiliation{Now at Institut d’Electronique et des Technologies du num\'eRique (IETR), UMR CNRS 6164, INSA Rennes, 35700 Rennes, France}
\email[]{jorge.ruiz-garcia@insa-rennes.fr}
\affiliation{Department of Electrical Engineering and Computer Science, The University of Michigan, Ann Arbor, 48109 USA}
\author{Anthony Grbic}
\email[]{agrbic@umich.edu}
%\thanks{}
\affiliation{Department of Electrical Engineering and Computer Science, The University of Michigan, Ann Arbor, 48109 USA}


\date{\today}

\begin{abstract}
Fully harnessing the vast design space enabled by metamaterials to control electromagnetic (EM) fields remains an open problem for researchers. Inverse-design techniques have shown to best exploit the degrees of freedom available in design, resulting in high-performing systems for wireless communications, sensing and analog signal processing. Nonetheless, fundamental yet powerful properties of metamaterials are still to be revealed. In this paper, we introduce the concept of Perfectly Matched Metamaterials (PMMs). PMMs are passive, inhomogeneous media that perform purely-refractive field transformations under different excitations. Their advantage lies in their simplicity, reflectionless behavior and suitability for both analytical and numerical design methods. Unlike Transformation Optics, PMM-based designs are devoid of coordinate transformations. Anisotropic unit cells are configured to control EM fields in a true-time delay manner. Simple analytical designs are reported which demonstrate the broadband capability of PMM devices. Proposed PMMs may find application in wideband beamforming and analog computing, realizing functionalities such as spatial filtering and signal pre-processing.
\end{abstract}

%\keywords{}

\maketitle

\section{Introduction\label{sec:Intro}}
Controlling electromagnetic (EM) fields remains a central challenge in wave physics. To this end, metamaterials have become a foundational platform. Metamaterials are subwavelength-textured structures capable of engineering EM field responses beyond natural limits. Nowadays, a new cohort of radiating, beamforming and analog computing metamaterials increasingly fulfill needs in communications and sensing. We have already witnessed basic wavefront manipulation through graded-index (GRIN) lenses \cite{Mrnka:2024,Bilitos:2024,Darabi:2018,Babayi:2019}. Further, more advance field control has been demonstrated by independently tailoring power and phase progression \cite{Martini:2015,Gashi:2025}. However, a growing need to perform more challenging functionalities is upon us. For instance, analog signal processing usually involves complicated EM field transformations \cite{Nikkhah:2024,Cordaro:2023,Li:2023_imaging,Zangeneh:2021,Hughes:2019}. These field transformations are generally achieved through the inverse-design of material parameters \cite{Ma:2024_invDesign,Li:2022_inverseDesign,Pestourie:2018}. Unlike analytical techniques, inverse design can better exploit a metamaterial's increased design space. Whereas researchers have focused on developing advanced and exotic field-controlling devices, few have paid attention to the bandwidth of such devices. 

Typically, intricate field transformations imply narrowband performance. There are two main reasons for this. One is the reflective nature of many inverse-designed structures \cite{Nikkhah:2024,Cordaro:2023,Nikkhah:2023_APS,Szymanski:2022,Hughes:2019}. These structures rely on internal reflections to realize a prescribed function. The other reason is the inherently frequency-dispersive unit cells employed in discretized realizations, like split-ring resonators \cite{Economou:2008,Pendry:1999}. Ultimately, the scientific community is in need of a design strategy to perform broadband, arbitrary field transformations. This is achieved by independently controlling the magnitude and phase of EM fields within a reflectionless environment. 

Transformation Optics (TO) has been commonly employed to satisfy these requirements. In 2006, Leonhardt \cite{Leonhardt:2006_TO} and Pendry, Schurig and Smith \cite{Pendry:2006_TO,Schurig:2006_TO} formalized and introduced TO as a design tool in EM and Optics. TO was promptly exploited in design by many researchers, such as Kwon and Werner \cite{Kwon:2008,Kwon:2009,Kwon:2010}. TO techniques together with metamaterials enabled unconventional field control with new cloaking, beam-controlling and antenna devices \cite{Kundtz:2011,Kwon:2010,Kwon:2014_bookChapter,McCall:2018}. TO rests on the invariability of Maxwell's equations under the expansion, compression and rotation of space, or equivalently under coordinate transformations. This implies that, to manipulate fields at will, one can just spatially modify the underlying permittivity $\boldsymbol{\overline{\overline{\varepsilon}}}$ and permeability $\muTensor$ of a medium. Under TO fundamentals, a prescribed field transformation can be performed that independently controls magnitude and phase. In addition, TO generally ensures reflectionless field transmission between the transformed and surrounding space. Consequently, TO tools have since been considered a panacea for achieving reflectionless wave manipulation.

However, TO techniques have notable drawbacks. For example, some basic TO structures are not actually reflectionless. In 2D environments, transforming a curved boundary into a straight boundary results in reflections between the transformed and original spaces \cite{Lin:2008_TO,Rahm:2008,Kundtz:2011,Kwon:2014_bookChapter}. Basic TO designs such as beam-expanders, compressors and lenses exhibit reflections. Additionally, TO generates spatially-variant, anisotropic media with both tensor electric and magnetic responses. Realizing such media is challenging. Optimization and simplification methods emerged to make TO devices realizable. In short, these methods: 1$)$ render the material parameters of transformed spaces purely scalar and non-magnetic (i.e., $\mu=1$), 2$)$ minimize possible reflections between transformed and original spaces. Conformal TO \cite{Leonhardt:2006_TO} and quasi-conformal TO \cite{Li_Pendry:2008_QCTO,Kwon:2012_QCTO} are the most representative examples. After applying these methods, the resulting structures can be straightforwardly implemented with dielectric materials. Of course, this is at the expense of the degrees of freedom offered by anisotropic media. Some of these methods rely on complicated, non-linear space transformations \cite{Kwon:2009}. Moreover, some space transformations must be solved numerically or map to nonphysical material parameters (e.g., infinite values) \cite{Kwon:2008}. In any case, these mappings are exclusively related to a specific field transformation, and must be accordingly redefined when changing the functionality. Finally, despite their complexity, the aforementioned methods do not always guarantee reflectionless field transformations \cite{Kwon:2010, Rahm:2008_finiteEmbed}.

\begin{figure}[t]
	\centering
	\includegraphics[width=1\columnwidth]{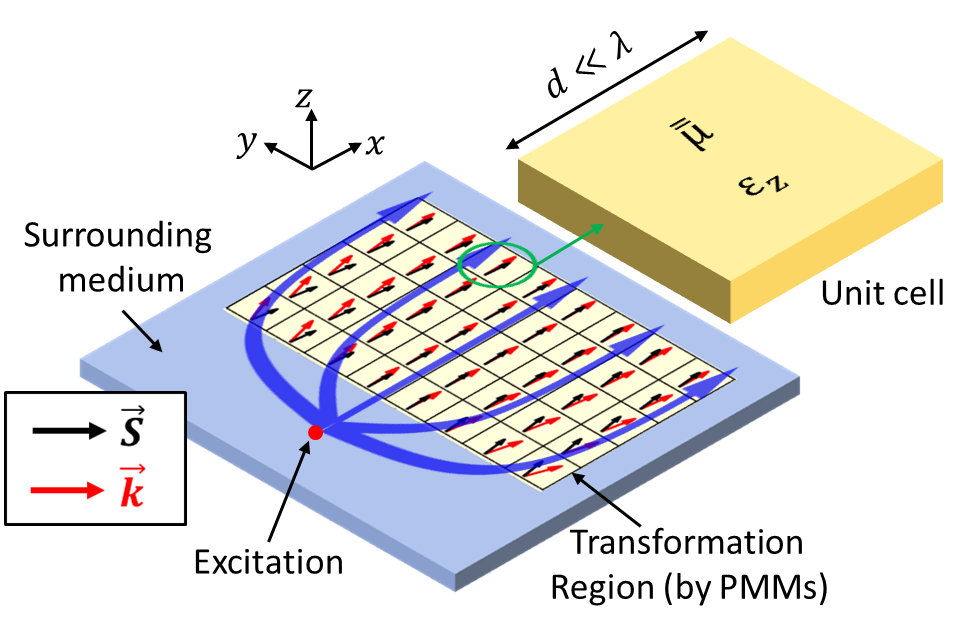}
	\caption{Representation of reflectionless field transformation (beam-collimator) using Perfectly Matched Metamaterials (PMMs). Blue lines symbolize the purely-refractive field progression, which combines with the independent control of Poynting vector $\boldsymbol{S}$ (power) and wavevector $\boldsymbol{k}$ (phase). The transformation region is formed from anisotropic unit cells with the material parameters given in \eqref{eq:pm_media_genEpsMu}. Unit cells have a size $d \ll \lambda$, where $\lambda$ is the free-space wavelength.}
	\label{fig:aniMedia_metamaterialPowerPhaseConcept}
\end{figure}

In this paper, we introduce a design approach to performing arbitrary reflectionless field transformations, overcoming many issues associated with TO. Consider a 2D inhomogeneous medium composed of anisotropic, homogeneous, subwavelength unit cells. Now, for a specified excitation, stipulate a desired field profile at the output of a given transformation region in space. This information is sufficient to analytically calculate the material parameters (scalar permittivity $\varepsilon$ and tensor permeability $\muTensor$) required in each unit cell to carry out the prescribed field transformation. The material parameters ensure that the unit cells are impedance-matched to each other for any direction of propagation. In other words, propagation within the metamaterial is all-angle reflectionless. Finally, the unit cells are also perfectly matched (impedance-matched for any angle of incidence) to the surrounding medium. Therefore, transmission across the input and output boundaries of the metamaterial is free of reflections as well. For all these reasons, we denote this new category of metamaterials as Perfectly Matched Metamaterials (PMMs). In short, PMMs can perform arbitrary field transformations relying on purely refractive wave propagation, much like true-time delay structures. This behavior is illustrated in Fig.~\ref{fig:aniMedia_metamaterialPowerPhaseConcept}.

PMMs combine the concept of perfectly matched media \cite{Gok:2016} with the field control offered by anisotropic metamaterials \cite{Gok:2013_Tailoring}. We will show that this enables reflectionless environments that provide independent control of power and phase progression. We first describe how material parameters in lossless, magnetically anisotropic, multi-layered media can be engineered to support reflectionless field transmission. Perfectly matched media can be interpreted as a stretched and rotated reference, isotropic medium. We describe how to associate power flow direction and phase progression with the material parameters of anisotropic metamaterials' unit cells. Enforcing perfect-matching conditions on these unit cells results in two degrees of freedom that enable arbitrary reflectionless field control. We show examples performing reflectionless beam collimation, which are not possible with TO methods. Design examples show an operating bandwidth from 5 to 30 GHz, resulting in approximately $140\%$ fractional bandwidth. These examples also confirm an inherent property of PMMs: they remain impedance-matched for any excitation, not only for the one they were designed for. The reported PMMs may pave the way to develop new broadband multifunctional devices.

In order to highlight PMMs' benefits, we establish a direct comparison with TO techniques and structures. Moreover, the name and properties of the proposed metamaterials may evoke the concept of Perfectly Matched Layers (PMLs) \cite{Berenger:1994,Sacks:1995_PML}. For this reason, we also comment on the main similarities and differences between PMMs and PMLs.

\section{The Concept of Perfect Matching}
\subsection{Analysis of magnetically anisotropic media \label{sec:MagnetAnisMedia}}
\begin{figure}[t]
	\centering
	\includegraphics[width=1\columnwidth]{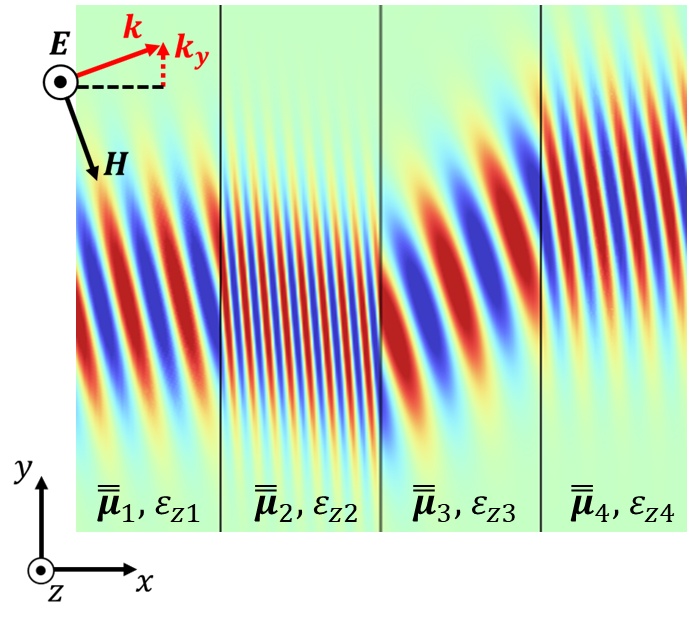}
	\caption{Time snapshot of $E_z$ for a plane wave propagating across multiple perfectly matched media. The material parameters for each medium are provided in Table~\ref{table:mu_eps_multilayer}. Note that each medium is reciprocal, i.e., $\mu_{xy}=\mu_{yx}$.}
	\label{fig:perfectMatch_media_Nlayers}
\end{figure}

Let us restrict ourselves to lossless, homogeneous, magnetically anisotropic media with relative permittivity $\boldsymbol{\overline{\overline{\varepsilon}}}$ and permeability $\boldsymbol{\overline{\overline{\mu}}}$ given by,
\begin{equation}
	\boldsymbol{\overline{\overline{\mu}}} = 
	\begin{bmatrix}
		\mu_{xx} &\mu_{xy} &0 \\ \mu_{yx} &\mu_{yy} &0 \\ 0 &0 &1 \\ 
	\end{bmatrix}, \quad
	\boldsymbol{\overline{\overline{\varepsilon}}} = 
	\begin{bmatrix}
		1 &0 &0 \\ 0 &1 &0 \\ 0 &0 &\varepsilon_z \\ 
	\end{bmatrix}.
	\label{eq:pm_media_genEpsMu}
\end{equation}
We will consider S-polarized (or $z$-polarized TEM) plane waves propagating in the $xy$-plane, as shown in Fig.~\ref{fig:perfectMatch_media_Nlayers}. The wavenumber is of the form $\boldsymbol{k}=(k_x,k_y,0)$, while the vector fields are $\boldsymbol{E}=(0,0,E_z)$ and $\boldsymbol{H}=(H_x,H_y,0)$ with time-harmonic dependence $e^{j\omega t}$. The differential form of Faraday's and Ampere's law are, 
\begin{equation}
	\nabla \times \boldsymbol{E} = -j \omega (\mu_0 \boldsymbol{\overline{\overline{\mu}}}) \cdot \boldsymbol{H}, \quad
	\nabla \times \boldsymbol{H} = j \omega (\varepsilon_0 \boldsymbol{\overline{\overline{\varepsilon}}}) \cdot \boldsymbol{E}.
	\label{eq:pm_media_maxwellEqs_diffForm}
\end{equation}
For the polarization and propagation of interest, we can write,
\begin{equation}
	%	\begin{split}
		-j(\boldsymbol{k} \times \boldsymbol{E}) = -j \omega \mu_0 \boldsymbol{\overline{\overline{\mu}}} \cdot \boldsymbol{H}, \quad
		-j(\boldsymbol{k} \times \boldsymbol{H}) = j \omega \varepsilon_0 \varepsilon_z \boldsymbol{E}.
		%	\end{split}
	\label{eq:pm_media_maxwellEqs}
\end{equation}
Performing the cross and dot products in \eqref{eq:pm_media_maxwellEqs} leads to,
\begin{equation}
	\begin{bmatrix}
		k_y E_z \\ -k_x E_z \\
	\end{bmatrix}
	= \omega \mu_0 
	\begin{bmatrix}
		\mu_{xx} &\mu_{xy} \\ \mu_{yx} &\mu_{yy} \\
	\end{bmatrix}
	\begin{bmatrix}
		H_x \\ H_y \\
	\end{bmatrix},
	\label{eq:pm_media_maxwellEqs_E_k}
\end{equation}
\begin{equation}
	k_y H_x - k_x H_y = \omega \varepsilon_0 \varepsilon_z E_z,
	\label{eq:pm_media_maxwellEqs_H_k}
\end{equation}
where we have used matrix notation. Let us multiply \eqref{eq:pm_media_maxwellEqs_E_k} and \eqref{eq:pm_media_maxwellEqs_H_k} by $1/E_z$, and use the relations $\omega \mu_0 = k_0 \eta_0$, $\omega \epsilon_0 = k_0/\eta_0$ ($k_0$ and $\eta_0$ are the free-space wavenumber and wave impedance, respectively) to obtain \cite{Gok:2016,Gok:2013_Tailoring},
\begin{equation}
	\begin{bmatrix}
		\overline{k}_y \\ \overline{k}_x \\
	\end{bmatrix}
	= 
	\begin{bmatrix}
		\mu_{xx} &-\mu_{xy} \\ -\mu_{yx} &\mu_{yy} \\
	\end{bmatrix}
	\begin{bmatrix}
		1/\overline{\eta}_y \\ 1/\overline{\eta}_x \\
	\end{bmatrix},
	\label{eq:pm_media_kEta_mu}
\end{equation}
\begin{equation}
	\frac{\overline{k}_x}{\overline{\eta}_x} + \frac{\overline{k}_y}{\overline{\eta}_y} = \varepsilon_z.
	\label{eq:pm_media_kEta_eps}
\end{equation}
In \eqref{eq:pm_media_kEta_mu} and \eqref{eq:pm_media_kEta_eps}, we have used the normalized wave impedances given by,
\begin{equation}
	\eta_x = -\frac{E_z}{H_y \eta_0}, \quad \eta_y = \frac{E_z}{H_x \eta_0},
	\label{eq:pm_media_waveImpedance}
\end{equation}
and normalized wavenumbers $\overline{k}_{x,y}=k_{x,y}/k_0$. Equations \eqref{eq:pm_media_kEta_mu} and \eqref{eq:pm_media_kEta_eps} relate wavevectors and wave impedances in a magnetically anisotropic medium. Both the wavevector and wave impedances are related through the material parameters of the medium.

Next, we will derive the dispersion equation for this type of medium. By substituting $E_z$ from \eqref{eq:pm_media_maxwellEqs_H_k} into \eqref{eq:pm_media_maxwellEqs_E_k} and some simple manipulations, we obtain the following expression,
\begin{equation}
	\left(
	\frac{1}{\varepsilon_z}
	\begin{bmatrix}
		\kyNorm^2 &-\kxNorm \kyNorm \\ -\kxNorm \kyNorm &\kxNorm^2 \\
	\end{bmatrix}
	-
	\begin{bmatrix}
		\mu_{xx} &\mu_{xy} \\ \mu_{yx} &\mu_{yy} \\
	\end{bmatrix}
	\right)
	\begin{bmatrix}
		H_x \\ H_y \\
	\end{bmatrix}
	= 
	0.
	\label{eq:pm_media_dispEq_1}
\end{equation}
For a given wavevector $\boldsymbol{k}$, the eigenfrequency $\omega$ can be obtained by solving the eigenvalue problem stated in \eqref{eq:pm_media_dispEq_1}. For a nontrivial solution of $\omega$, the determinant of the coefficient matrix must vanish. Setting the determinant equal to zero yields the dispersion equation for the magnetically anisotropic medium,
\begin{equation}
	\begin{split}
		\left( \frac{\overline{k}_y^2}{\varepsilon_z} - \mu_{xx} \right)& 
		\left( \frac{\overline{k}_x^2}{\varepsilon_z} - \mu_{yy} \right) - \\
		&\left( \frac{\overline{k}_x \overline{k}_y}{\varepsilon_z} + \mu_{xy} \right)
		\left( \frac{\overline{k}_x \overline{k}_y}{\varepsilon_z} + \mu_{yx} \right)
		= 0.
	\end{split}
	\label{eq:pm_media_dispEq_2}
\end{equation}
By introducing the determinant of the permeability tensor $|\boldsymbol{\overline{\overline{\mu}}}|$, the dispersion equation can be alternatively written as follows,
\begin{equation}
	\overline{k}_x^2 \mu_{xx} + \overline{k}_y^2 \mu_{yy} +
	\overline{k}_x \overline{k}_y \left(\mu_{xy} + \mu_{yx}\right)
	= \varepsilon_z |\boldsymbol{\overline{\overline{\mu}}}|.
	\label{eq:gen_dispEq_magAnisoMedia}
\end{equation}

\begin{table}[t]%The best place to locate the table environment is directly after its first reference in text
	\caption{\label{table:mu_eps_multilayer}%
		Material parameters for the multilayered medium shown \figref{fig:perfectMatch_media_Nlayers}.
	}
	\begin{ruledtabular}
		\begin{tabular}{ccccc}
			\textrm{Medium}&
			\textrm{$\mu_{xx}$}&
			\textrm{$\mu_{xy}$=$\mu_{yx}$}&
			\textrm{$\mu_{yy}$}&
			\textrm{$\varepsilon_{z}$}\\
			\colrule
			1 & 1 & 0 & 1 & 1\\
			2 & 0.377 & -0.339 & 2.958 & 2.653\\
			3 & 1.140 & 1.450 & 2.721 & 0.877\\
			4 & 0.667 & 0 & 1.5 & 1.5\\
		\end{tabular}
	\end{ruledtabular}
\end{table}

\subsection{All-angle reflectionless condition \label{sec:AllAngleConditions}}
It is possible to ensure perfect impedance matching between different media of the type described in Section~\ref{sec:MagnetAnisMedia}, separated by a planar boundary along the $y$-axis (see Fig.~\ref{fig:perfectMatch_media_Nlayers}). In other words, one can realize reflectionless transmission across the boundary regardless of the wave's propagation direction. Hereinafter, this effect will be referred to as all-angle reflectionless transmission, or perfect matching \cite{Gok:2016}. Perfect matching across a boundary separating two media requires that the normal wave impedance $\eta_x$ is the same for both media for any value of the tangential wavenumber $k_y$. For this reason, we need to find a relation between $\eta_x$ and $k_y$. Such a relation will allow us to find the form of the material parameters needed to achieve perfect matching.

First, we will use the system of equations in \eqref{eq:pm_media_kEta_mu} to solve for $\overline{\eta}_x$. From the first equation we have, 
\begin{equation}
	\frac{1}{\overline{\eta}_y} = \frac{1}{\mu_{xx}} \left(\overline{k}_y + \frac{\mu_{xy}}{\overline{\eta}_x}\right).
	\label{eq:pm_etay_interms_kx_ky_etax}
\end{equation}
Substituting $1/\overline{\eta}_y$ into the second equation yields,
\begin{equation}
	\frac{1}{\overline{\eta}_x} = \frac{1}{|\boldsymbol{\overline{\overline{\mu}}}|} \left( \overline{k}_x \mu_{xx} + \overline{k}_y \mu_{yx} \right).
	\label{eq:pm_etax_interms_kx_ky}
\end{equation}
Using \eqref{eq:pm_etax_interms_kx_ky}, we can express $\overline{k}_x$ in terms of $\overline{\eta}_x$ and $\overline{k}_y$,
\begin{equation}
	\overline{k}_x = \frac{1}{\mu_{xx}} \left( \frac{|\boldsymbol{\overline{\overline{\mu}}}|}{\overline{\eta}_x} 
	- \overline{k}_y \mu_{yx}	\right).
	\label{eq:pm_kx_interms_etax_ky}
\end{equation}
Substituting the expression above into the dispersion equation given by \eqref{eq:gen_dispEq_magAnisoMedia} yields the following relation,
\begin{equation}
	|\boldsymbol{\overline{\overline{\mu}}}| \left( \frac{1}{\overline{\eta}_x} \right)^2 + 
	\frac{\overline{k}_y}{\overline{\eta}_x} \left(\mu_{xy} - \mu_{yx}\right) + 
	\overline{k}_y^2
	= \mu_{xx} \varepsilon_z.
	\label{eq:pm_etax_ky_general}
\end{equation}
Equation \eqref{eq:pm_etax_ky_general} is satisfied by all S-polarized plane waves in the magnetically, anisotropic medium. In addition, \eqref{eq:pm_etax_ky_general} relates $\overline{\eta}_x$ and $\overline{k}_y$ through the material parameters of the medium. By denoting $\Delta=|\muTensor|$ and reorganizing the terms, the relation in \eqref{eq:pm_etax_ky_general} can be rewritten as,
\begin{equation}
	\left( \frac{1}{\overline{\eta}_x} \right)^2 \left(\frac{\Delta}{\varepsilon_z \mu_{xx}}\right) + 
	\frac{\overline{k}_y}{\overline{\eta}_x} \left( \frac{\mu_{xy} - \mu_{yx}}{\varepsilon_z \mu_{xx}} \right) +
	\overline{k}_y^2 \left(\frac{1}{\varepsilon_z \mu_{xx}}\right) = 1.
	\label{eq:pm_etax_ky_general_modif}
\end{equation}

Let us apply now the conditions of perfect matching. Medium 1 and medium 2 (shown in Fig.~\ref{fig:perfectMatch_media_Nlayers})  must have the same normal wave impedance, that is $\overline{\eta}_{x1}=\overline{\eta}_{x2}$, for any $k_y$. In addition, phase-matching ensures that the tangential wavenumber is continuous across the boundary, $\overline{k}_{y1}=\overline{k}_{y2}$. Since \eqref{eq:pm_etax_ky_general_modif} must be satisfied in medium 1 and medium 2, we can write,
\begin{equation}
	\begin{split}
		&\left( \frac{1}{\overline{\eta}_x} \right)^2 
		\left[\frac{\Delta_1}{(\varepsilon_{z} \mu_{xx})_1} - \frac{\Delta_2}{(\varepsilon_{z} \mu_{xx})_2}\right] + \\
		&\frac{\overline{k}_y}{\overline{\eta}_x} \left[ \frac{(\mu_{xy} - \mu_{yx})_1}{(\varepsilon_{z} \mu_{xx})_1} - 
		\frac{(\mu_{xy} - \mu_{yx})_2}{(\varepsilon_{z} \mu_{xx})_2} \right] + \\
		&\overline{k}_y^2 \left[\frac{1}{(\varepsilon_{z} \mu_{xx})_1} - \frac{1}{(\varepsilon_{z} \mu_{xx})_2}\right] = 0,
	\end{split}
	\label{eq:pm_etax_ky_general_2media}
\end{equation}
where the subscripts refer to the medium. For \eqref{eq:pm_etax_ky_general_2media} to be true for any combination of $\overline{\eta}_x$ and $\overline{k}_y$, the following conditions must hold,
\begin{gather}
	(\varepsilon_{z} \mu_{xx})_1 = (\varepsilon_{z} \mu_{xx})_2, \quad
	\Delta_1 = \Delta_2, \nonumber \\
	(\mu_{xy} - \mu_{yx})_1 = (\mu_{xy} - \mu_{yx})_2.
	\label{eq:pm_pefectMatching_conditions}
\end{gather}
These conditions can be generalized for multilayered media containing any number of layers separated by planar boundaries. This is illustrated in \figref{fig:perfectMatch_media_Nlayers}. Defining the normal refractive index $n_{yy}^2=\varepsilon_z \mu_{xx}$ and the constant $\Lambda_{xy}=\mu_{xy} - \mu_{yx}$, all the layers will be perfectly matched as long as each layer $i$ satisfies the following conditions,
\begin{equation}
	\left(\varepsilon_{z} \mu_{xx}\right)_i = n_{yy}^2, \quad
	\Delta_i = \Delta, \quad
	\left(\mu_{xy} - \mu_{yx}\right)_i = \Lambda_{xy}.
	\label{eq:pm_pefectMatching_conditions_Nlayers}
\end{equation}
This means that the material parameters $\varepsilon_{z}$ and $\boldsymbol{\overline{\overline{\mu}}}$ in all the different layers are dictated by the all-angle reflectionless constants $n_{yy}^2$, $\Delta$ and $\Lambda_{xy}$. The material parameters used in the multilayered medium shown in \figref{fig:perfectMatch_media_Nlayers} are provided in Table~\ref{table:mu_eps_multilayer}. One can easily verify that all layers have $\Delta=n_{yy}^2=1$ and $\Lambda_{xy}=0$.

Perfect-matching conditions can be likewise derived for media separated by planar boundaries along the $x$-axis. This is described in Appendix~\ref{sec:appendix_PMM_xBoundary}. 

\subsection{Reciprocal media}
For the polarization of interest, the conditions given in \eqref{eq:pm_pefectMatching_conditions_Nlayers} are general and valid for any magnetically, anisotropic medium. In the particular case of reciprocal media, the cross-diagonal entries of $\boldsymbol{\overline{\overline{\mu}}}$ are the same, $\mu_{xy}=\mu_{yx}$. Then, the formulation in Section~\ref{sec:MagnetAnisMedia} and \ref{sec:AllAngleConditions} reduces to that presented in \cite{Gok:2016}. The dispersion equation given in \eqref{eq:gen_dispEq_magAnisoMedia} becomes,
\begin{equation}
	\overline{k}_x^2 \mu_{xx} + \overline{k}_y^2 \mu_{yy} +
	2 \overline{k}_x \overline{k}_y \mu_{xy}
	= \varepsilon_z |\boldsymbol{\overline{\overline{\mu}}}|.
	\label{eq:gen_dispEq_magAnisoMedia_reciproc}
\end{equation}
Furthermore, the quadratic equation \eqref{eq:pm_etax_ky_general_modif} relating $\overline{\eta}_x$ and $\overline{k}_y$ reduces to, 
\begin{equation}
	\left( \frac{1}{\overline{\eta}_x} \right)^2 \left(\frac{\Delta}{\varepsilon_z \mu_{xx}}\right) + 
	\overline{k}_y^2 \left(\frac{1}{\varepsilon_z \mu_{xx}}\right) = 1.
	\label{eq:pm_etax_ky_reciproc_modif}
\end{equation}
Finally, $\Lambda_{xy}$ vanishes and only the first two conditions given by \eqref{eq:pm_pefectMatching_conditions_Nlayers} remain,
\begin{equation}
	\left(\varepsilon_{z} \mu_{xx}\right)_i = n_{yy}^2, \quad
	\Delta_i = \Delta.
	\label{eq:pm_pefectMatching_conditions_Nlayers_reciproc}
\end{equation}
Using these two conditions, we can write a general form of the material parameters that ensures perfect matching,
\begin{equation}
	\boldsymbol{\overline{\overline{\mu}}} = 
	\begin{bmatrix}
		\mu_{xx} &\mu_{xy} \\ \mu_{xy} &\frac{\left(\Delta + \mu_{xy}^2\right)}{\mu_{xx}} \\
	\end{bmatrix}, \quad
	\varepsilon_z = \frac{n_{yy}^2}{\mu_{xx}}.
	\label{eq:pm_pefectMatching_generalForm_muEps}
\end{equation}
Equivalently, by denoting $A=1/\mu_{xx}$ and $B=\mu_{xy}$, the material parameters can be expressed as \cite{Gok:2016},
\begin{equation}
	\boldsymbol{\overline{\overline{\mu}}} = 
	\begin{bmatrix}
		1/A &B \\ B &A \left(\Delta + B^2\right) \\
	\end{bmatrix}, \quad
	\varepsilon_z = n_{yy}^2 A.
	\label{eq:pm_pefectMatching_ABform_muEps}
\end{equation}
Therefore, perfect impedance matching reduces the degrees of freedom for reciprocal, magnetically anisotropic media from four ($\mu_{xx}$, $\mu_{xy}$, $\mu_{yy}$, $\varepsilon_{z}$) to two ($\mu_{xx}$, $\mu_{xy}$, or $A$, $B$).

A more physical connection can be established between the two available degrees of freedom and the properties of perfectly matched media. The permeability tensor in \eqref{eq:pm_media_genEpsMu} can be rewritten in terms of its eigenvalue $C$ and the tensor's rotation angle $\Psi$ as,
\begin{equation}
	\boldsymbol{\overline{\overline{\mu}}} = R^T(\Psi) \left(
	\sqrt{\Delta}
	\begin{bmatrix}
		C &0 \\ 0 &1/C \\
	\end{bmatrix} \right) R(\Psi),
	\label{eq:pm_pefectMatching_CpsiForm_mu}
\end{equation}
with,
\begin{equation}
	R(\Psi) = 
	\begin{bmatrix}
		\cos\Psi &-\sin\Psi \\ \sin\Psi &\cos\Psi \\
	\end{bmatrix}.
	\label{eq:pm_rotationMatrix}
\end{equation}
Using \eqref{eq:pm_rotationMatrix} in \eqref{eq:pm_pefectMatching_CpsiForm_mu} yields,
\begin{equation}
	\boldsymbol{\overline{\overline{\mu}}} = 
	\frac{\sqrt{\Delta}}{C}
	\begin{bmatrix}
		C^2 \cos^2\Psi + \sin^2\Psi &(1-C^2)\sin\Psi \cos\Psi
		\\[6pt] (1-C^2)\sin\Psi \cos\Psi &\cos^2\Psi + C^2\sin^2\Psi \\
	\end{bmatrix}.
	\label{eq:pm_pefectMatching_mu_CpsiDelta}
\end{equation}
This representation can be interpreted as a stretched and rotated version of an isotropic medium with relative permeability $\sqrt{\Delta}$. It is clear that the determinant of $\boldsymbol{\overline{\overline{\mu}}}$ in \eqref{eq:pm_pefectMatching_mu_CpsiDelta} is $\Delta$. By substituting the entry $\mu_{xx}$ from \eqref{eq:pm_pefectMatching_mu_CpsiDelta} into the condition $\varepsilon_{z} \mu_{xx} = n_{yy}^2$, we obtain,
\begin{equation}
	\varepsilon_{z} = \frac{C}{\sqrt{\Delta}} \left( \frac{n_{yy}^2}{C^2 \cos^2\Psi + \sin^2\Psi} \right).
	\label{eq:pm_pefectMatching_CpsiForm_eps}
\end{equation}
Expressions \eqref{eq:pm_pefectMatching_mu_CpsiDelta} and \eqref{eq:pm_pefectMatching_CpsiForm_eps} provide material parameters that satisfy the all-angle reflectionless conditions across $y$-directed boundaries given in \eqref{eq:pm_pefectMatching_conditions_Nlayers_reciproc}. Consequently, perfectly matched media can be engineered through the stretching factor $C$ and rotation $\Psi$. The immediate question that emerges is, what field control do these two degrees of freedom provide? This is described in the next section.

\section{Field Control with \\ Perfectly Matched Metamaterials}
In this section, we show that the PMMs' two degrees of freedom enable reflectionless field transformations with independent control of power and phase progression. For this purpose, we resort to metamaterials formed from homogeneous, magnetically anisotropic unit cells of dimension $d$, as shown in Fig.~\ref{fig:aniMedia_metamaterialPowerPhaseConcept}. To realize a prescribed functionality, the local power flow direction and phase progression can be manipulated via the material parameters of each unit cell \cite{Gok:2013_Tailoring}. In addition, we can realize reflectionless field transformations through the perfect matching condition. To summarize, it will be shown that power flow and phase can be independently controlled while maintaining perfect matching. Reciprocal media will be considered.

\begin{figure}[t]
	\centering
	\includegraphics[width=0.6\columnwidth]{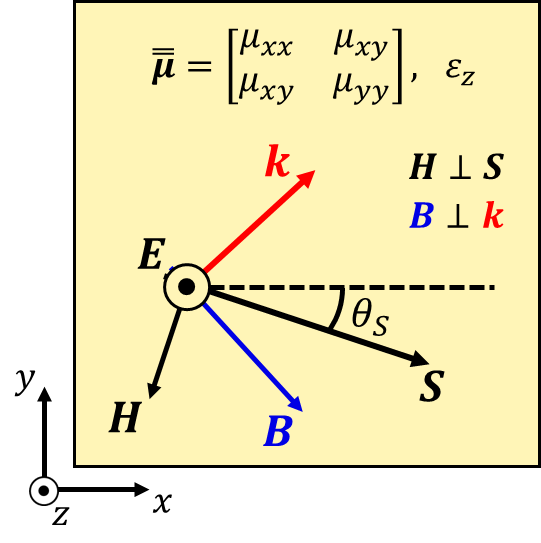}
	\caption{Representation of fields, wavevector $\boldsymbol{k}$ and Poynting vector $\boldsymbol{S}$ in a magnetically, anisotropic unit cell. Since the permeability is a tensor, the magnetic flux density $\boldsymbol{B}$ and magnetic field intensity $\boldsymbol{H}$ may not be parallel.}
	\label{fig:aniMedia_kS_nonAligned}
\end{figure}

Since the unit cells are subwavelength ($d \ll \lambda$, where $\lambda$ is the free-space wavelength), a local plane wave approximation can be applied. This allows the assumption that each cell locally controls the power flow direction and phase velocity of a planar wavefront. A study of crystal optics tells us that anisotropic media can support phase and group velocities that are non-aligned \cite{Kong:1986}. In magnetically anisotropic media, this is due to the fact that the magnetic flux density $\boldsymbol{B}$ is not parallel to the magnetic field intensity $\boldsymbol{H}$. As a result, the Poynting vector $\boldsymbol{S}$ and wavevector $\boldsymbol{k}$ may not be parallel, as shown in Fig.~\ref{fig:aniMedia_kS_nonAligned}. We will show how to relate $\boldsymbol{k}$ and the direction of $\boldsymbol{S}$ to $\varepsilon_{z}$ and $\boldsymbol{\overline{\overline{\mu}}}$.

First, let us calculate the direction of $\boldsymbol{S}$. The complex power flow is represented by the Poynting vector $\boldsymbol{S}=\boldsymbol{E} \times \boldsymbol{H}^*$. For the polarization of interest, $\boldsymbol{S}$ can be expressed as,
\begin{equation}
	\boldsymbol{S} = \boldsymbol{E} \times \boldsymbol{H}^* = \left(-E_z H^*_y, E_z H^*_x\right).
	\label{eq:aniMedia_PoyntingVector}
\end{equation}
Multiplying the expression above by $E_z/E^*_z$ and using the wave impedances defined in \eqref{eq:pm_media_waveImpedance}, the Poynting vector can be expressed in terms of $\eta$ as,
\begin{equation}
	\boldsymbol{S} = E_z^2 \left(\frac{1}{\eta_x}, \frac{1}{\eta_y}\right).
	\label{eq:aniMedia_PoyntingVector_eta}
\end{equation}
Note that we are considering lossless media, where the wave impedances are real and therefore $\eta_x=\eta_x^*$ and $\eta_y=\eta_y^*$. We will denote $\theta_S$ as the angle between $\boldsymbol{S}$ and the $x$-axis (see Fig.~\ref{fig:aniMedia_kS_nonAligned}). This angle can be obtained from the components of $\boldsymbol{S}$ as $\tan \theta_S = (S_y/S_x)$, which according to \eqref{eq:aniMedia_PoyntingVector_eta} yields,
\begin{equation}
	\kappa = \tan \theta_S = \eta_x / \eta_y.
	\label{eq:aniMedia_PoyntingVector_eta_direction}
\end{equation}
Therefore, the direction of $\boldsymbol{S}$ is given by the ratio of wave impedances, denoted as $\kappa$.

Next, we will derive an expression for $\boldsymbol{\overline{\overline{\mu}}}$ that depends on the local power flow direction and wavevector. Two equations can be derived from \eqref{eq:pm_media_kEta_eps} by separately multiplying by $\overline{\eta}_x$ and $\overline{\eta}_y$, 
\begin{equation}
	\overline{\eta}_x \varepsilon_z = \overline{k}_x + \frac{\overline{\eta}_x}{\overline{\eta}_y} \overline{k}_y, \quad
	\overline{\eta}_y \varepsilon_z = \frac{\overline{\eta}_y}{\overline{\eta}_x} \overline{k}_x + \overline{k}_y.
	\label{eq:aniMedia_etaX_etaY_kappa1}
\end{equation}
Using the definition of $\kappa$, \eqref{eq:aniMedia_etaX_etaY_kappa1} can be rewritten as \cite{Gok:2013_Tailoring},
\begin{equation}
	\overline{\eta}_x = \frac{\overline{k}_x + \kappa \overline{k}_y}{\varepsilon_z}, \quad
	\overline{\eta}_y = \frac{\overline{k}_x/\kappa + \overline{k}_y}{\varepsilon_z}.
	\label{eq:aniMedia_etaX_etaY_kappa}
\end{equation}
Now, we solve for the diagonal entries of $\boldsymbol{\overline{\overline{\mu}}}$ in \eqref{eq:pm_media_kEta_mu}, yielding,
\begin{equation}
	\mu_{xx} = \etayNorm \left( \kyNorm + \frac{\mu_{xy}}{\etaxNorm} \right), \quad
	\mu_{yy} = \etaxNorm \left( \kxNorm + \frac{\mu_{xy}}{\etayNorm} \right).
	\label{eq:aniMedia_muxx_muyy}
\end{equation}
Substituting the wave impedances given by \eqref{eq:aniMedia_etaX_etaY_kappa} into \eqref{eq:aniMedia_muxx_muyy} the permeability tensor can be written as,
\begin{equation}
	\muTensor =
	\begin{bmatrix}
		\frac{\kyNorm}{\varepsilon_z} \left(\frac{\kxNorm}{\kappa} + \kyNorm\right) + \frac{\mu_{xy}}{\kappa} 
		&\mu_{xy} \\ \mu_{xy} 
		&\frac{\kxNorm}{\varepsilon_z} \left(\kxNorm + \kappa \kyNorm\right) +  \kappa \mu_{xy} \\
	\end{bmatrix}.
	\label{eq:aniMedia_muTensor_powerPhase_eps_muxy}
\end{equation}
The expression in \eqref{eq:aniMedia_muTensor_powerPhase_eps_muxy} describes $\muTensor$ in terms of the local power flow direction ($\kappa$) and phase progression ($\kxNorm, \kyNorm$), with two remaining degrees of freedom: $\varepsilon_{z}$ and $\mu_{xy}$. These two degrees of freedom will be used to satisfy the perfect matching condition.

Let us consider the structure described in Fig.~\ref{fig:aniMedia_metamaterialPowerPhaseConcept}. The transformation region, which transforms the field excited by the source, is formed from PMM unit cells and embedded into a homogeneous surrounding medium. The surrounding medium is assumed to be lossless, homogeneous and isotropic, with a relative permittivity $\varepsilon_{z,s}$ and relative permeability $\muTensor_s=\overline{\overline{I}} \sqrt{\Delta}$, where $\overline{\overline{I}}$ is the unit dyadic. All unit cells of the transformation region (see Fig.~\ref{fig:aniMedia_metamaterialPowerPhaseConcept}) are assumed to have a relative permeability of the form given by \eqref{eq:aniMedia_muTensor_powerPhase_eps_muxy}. If all unit cells satisfy the second condition in \eqref{eq:pm_pefectMatching_conditions_Nlayers_reciproc}, then $|\muTensor|=\Delta$ and from \eqref{eq:aniMedia_muTensor_powerPhase_eps_muxy} we obtain,
\begin{equation}
	\mu_{xy} = \frac{\varepsilon_z \kappa \Delta}{\left(\kxNorm + \kappa \kyNorm\right)^2} 
	- \frac{\kxNorm \kyNorm}{\varepsilon_{z}}.
	\label{eq:pmm_muxy}
\end{equation}
Substituting \eqref{eq:pmm_muxy} into \eqref{eq:aniMedia_muTensor_powerPhase_eps_muxy} yields,
\begin{equation}
	\muTensor =
	\begin{bmatrix}
		\frac{\varepsilon_{z} \Delta}{\left(\kxNorm + \kappa \kyNorm\right)^2} + \frac{\kyNorm^2}{\varepsilon_z} 
		&\frac{\varepsilon_z \kappa \Delta}{\left(\kxNorm + \kappa \kyNorm\right)^2} 
		- \frac{\kxNorm \kyNorm}{\varepsilon_{z}} \\[6pt] 
		\frac{\varepsilon_z \kappa \Delta}{\left(\kxNorm + \kappa \kyNorm\right)^2} 
		- \frac{\kxNorm \kyNorm}{\varepsilon_{z}} 
		&\frac{\varepsilon_{z} \Delta \kappa^2}{\left(\kxNorm + \kappa \kyNorm\right)^2} + \frac{\kxNorm^2}{\varepsilon_z} \\
	\end{bmatrix}.
	\label{eq:pmm_mu_1}
\end{equation}
At this point, $\varepsilon_z$ can be arbitrarily chosen. To minimize reflections between the unit cells (internal reflections) for a prescribed field transformation, $\varepsilon_{z}$ can be optimized \cite{Gok:2013_Tailoring}. However, it is also possible to derive a closed-form expression for $\varepsilon_{z}$. If the surrounding medium has a refractive index given by $n_{yy}=\sqrt{(\varepsilon_z \mu_{xx})_s}$, one can apply the first condition in \eqref{eq:pm_pefectMatching_conditions_Nlayers_reciproc}. Substituting the $\boldsymbol{\hat{x}\hat{x}}$ entry of the tensor given in \eqref{eq:pmm_mu_1} into the expression $\varepsilon_{z} \mu_{xx}=n_{yy}^2$ results in a quadratic equation for $\varepsilon_{z}$. Its solution is given by,
\begin{equation}
	\varepsilon_{z} = \pm \left(\kxNorm + \kappa \kyNorm\right) \sqrt{ \frac{\left(n_{yy}^2 - \kyNorm^2\right)}{\Delta} }.
	\label{eq:pmm_eps}
\end{equation}
Since positive permittivities are simpler to realize, we will retain the positive solution of $\varepsilon_{z}$. By substituting \eqref{eq:pmm_eps} into \eqref{eq:pmm_mu_1}, the closed-form expression for $\muTensor$ reduces to,
	\begin{equation}
		\muTensor =
		\frac{1}{\varepsilon_z}
		\begin{bmatrix}
			n_{yy}^2 
			&\kappa \left(n_{yy}^2 - \kyNorm^2\right) - \kxNorm \kyNorm \\[6pt] 
			\kappa \left(n_{yy}^2 - \kyNorm^2\right) - \kxNorm \kyNorm 
			&\kappa^2 \left(n_{yy}^2 - \kyNorm^2\right) + \kxNorm^2 \\
		\end{bmatrix}.
		\label{eq:pmm_mu}
	\end{equation}
One can easily verify that \eqref{eq:pmm_eps} and \eqref{eq:pmm_mu} satisfy the perfect matching conditions $|\boldsymbol{\overline{\overline{\mu}}}|=\Delta$ and $\varepsilon_z \mu_{xx} = n_{yy}^2$. We have just shown that, for a stipulated local power flow direction and phase progression, the required material parameters of the unit cells can be analytically calculated using \eqref{eq:pmm_eps} and \eqref{eq:pmm_mu}. These material parameters provide unit cells that are perfectly matched to each other, and to the surrounding medium. These fundamental characteristics (closed-form expressions for $\muTensor$ and $\varepsilon_z$, and inherently perfectly matched unit cells) define the concept of perfectly matched metamaterials. 

Expressions \eqref{eq:pmm_eps} and \eqref{eq:pmm_mu} ensure perfect matching across $y$-directed boundaries. The formulation for perfect matching across $x$-directed boundaries can be derived in a similar manner. 

\section{Design examples \label{sec:examples_ideal}}
In this section, we show some design examples in order to validate the theoretical concept of PMMs. All the presented structures are formed from unit cells whose material parameters ensure perfect matching across $y$-directed boundaries. The employed material parameters are non-dispersive. Specifically, we assume that local $\varepsilon_{z}$ and $\muTensor$ do not vary with frequency. PMMs enable true-time delay field transformations, ideally resulting in broadband capabilities. In practice, such broadband capabilities can be preserved by using very low-dispersive unit cells \cite{Ruiz:2024,Ruiz:2024_eucap}.

The design procedure involving PMMs is simple: i$)$ choose an excitation (point source, tapered waveguide, plane wave, etc.) and stipulate output field; ii$)$ obtain required local power flow and phase progression in each unit cell; iii$)$ calculate the corresponding $\varepsilon_{z}$ and $\muTensor$ in all cells using \eqref{eq:pmm_eps} and \eqref{eq:pmm_mu}, respectively. In addition to the examples presented in this section, more designs exhibiting broadband field transformations are included in the Supplemental Material \cite{Ruiz:2024_PMM_SupplMaterial}.

\begin{figure}[t]
	\centering
	\includegraphics[width=1\columnwidth]{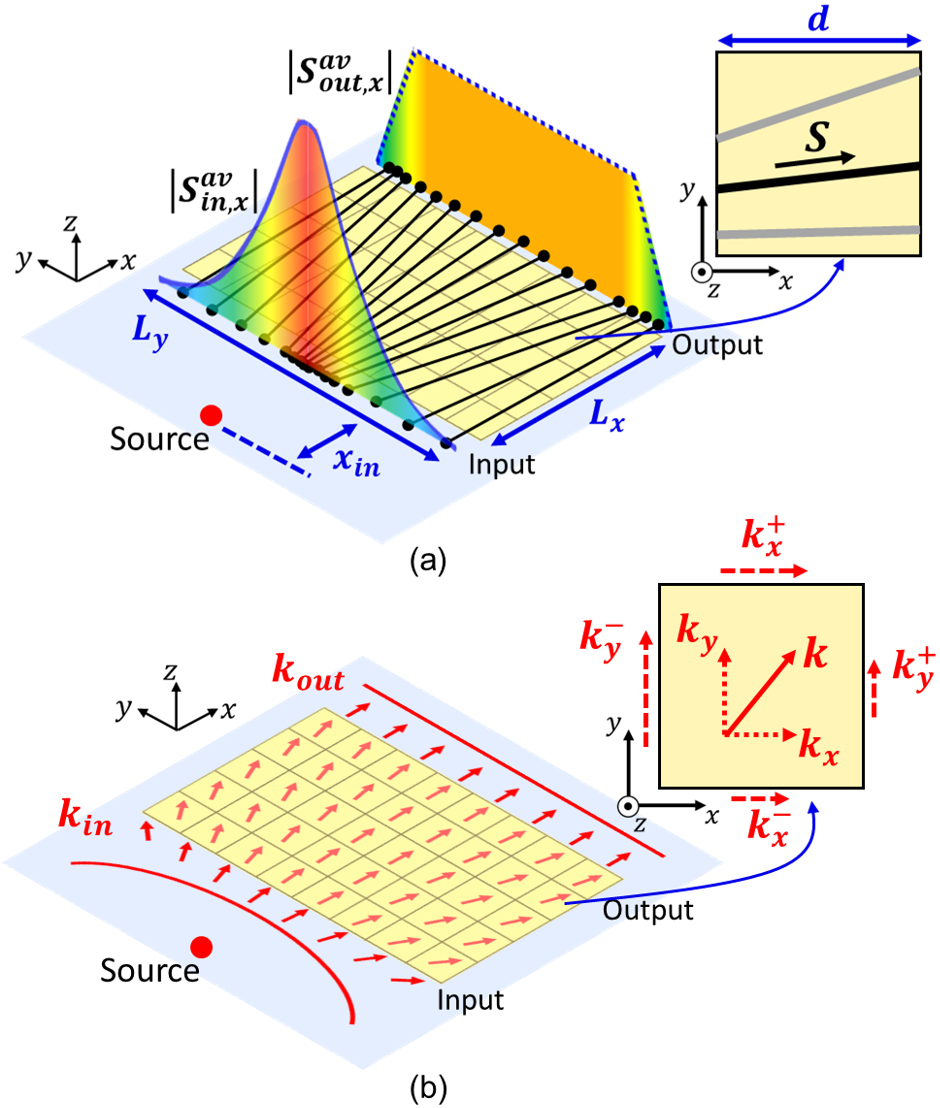}
	\caption{Representation of design procedure and beam-collimator performance. (a) The input and output power density profiles are discretized into the same number of power points. These points are connected through straight lines, whose direction dictates $\kappa$ in each unit cell. In the inset, the black line represents the closest line to the unit cell's center, while the gray lines represent other lines passing through the cell (see the Supplemental Material \cite{Ruiz:2024_PMM_SupplMaterial}). (b) The local phase progression follows a linear evolution from input to output along each $x$-directed row of cells. The inset shows the wavenumbers at the boundaries of each unit cell, whose average provides $k_x$ and $k_y$ (see the Supplemental Material \cite{Ruiz:2024_PMM_SupplMaterial}).}
	\label{fig:example_PMM_designConcept}
\end{figure}

\begin{figure}[t]
	\centering
	\subfloat[]{
		\centering
		\includegraphics[width=1\columnwidth]{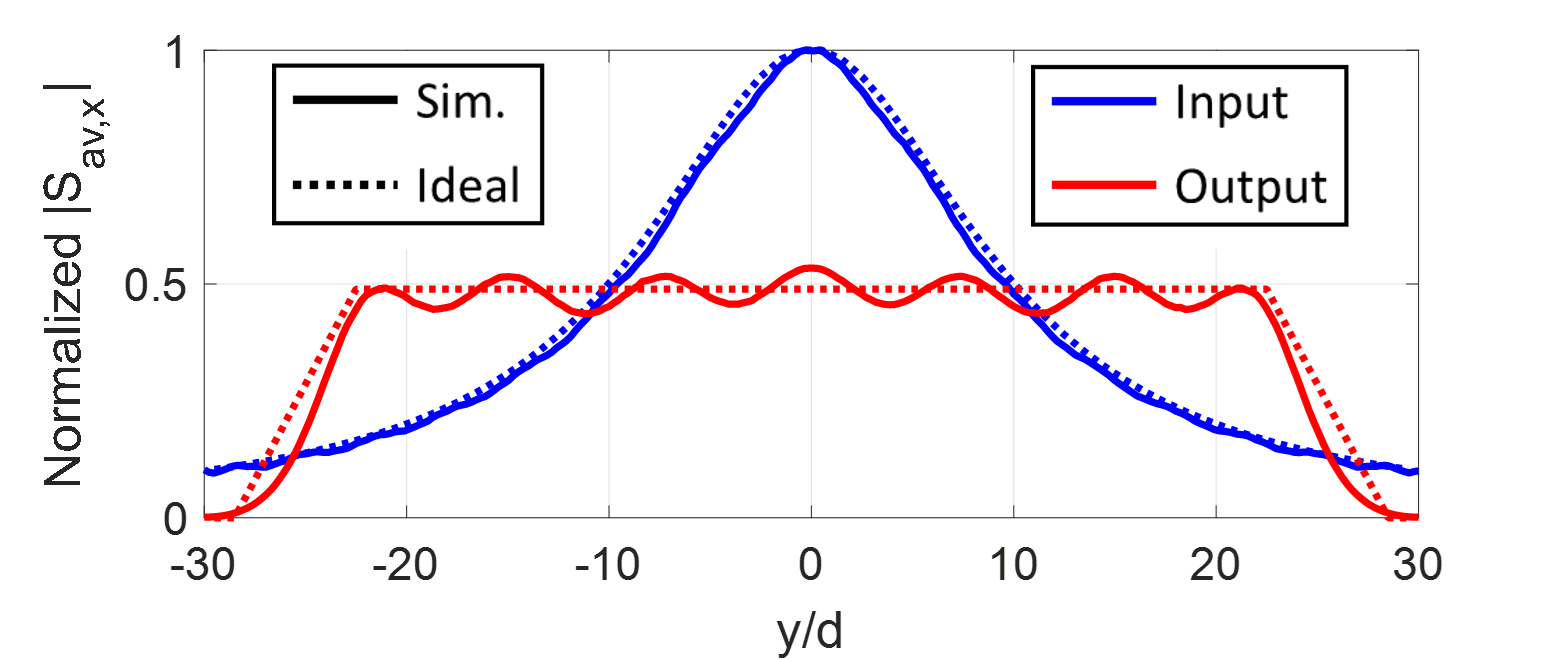}
		\label{fig:beamCollim_powerProfile1D}
	}
	\\
	\subfloat[]{
		\centering
		\includegraphics[width=1\columnwidth]{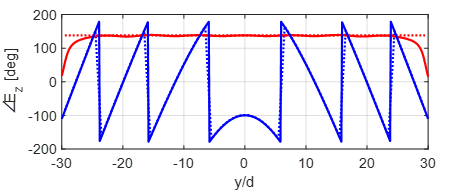}
		\label{fig:beamCollim_phaseProfile1D}
	}
	\caption{Desired and simulated power density and phase profiles at input and output interfaces of the transformation region at 10 GHz.}
	\label{fig:example_beamCollim_powerPhase}
\end{figure}

\begin{figure}[t]
	\centering
	\includegraphics[width=1\columnwidth]{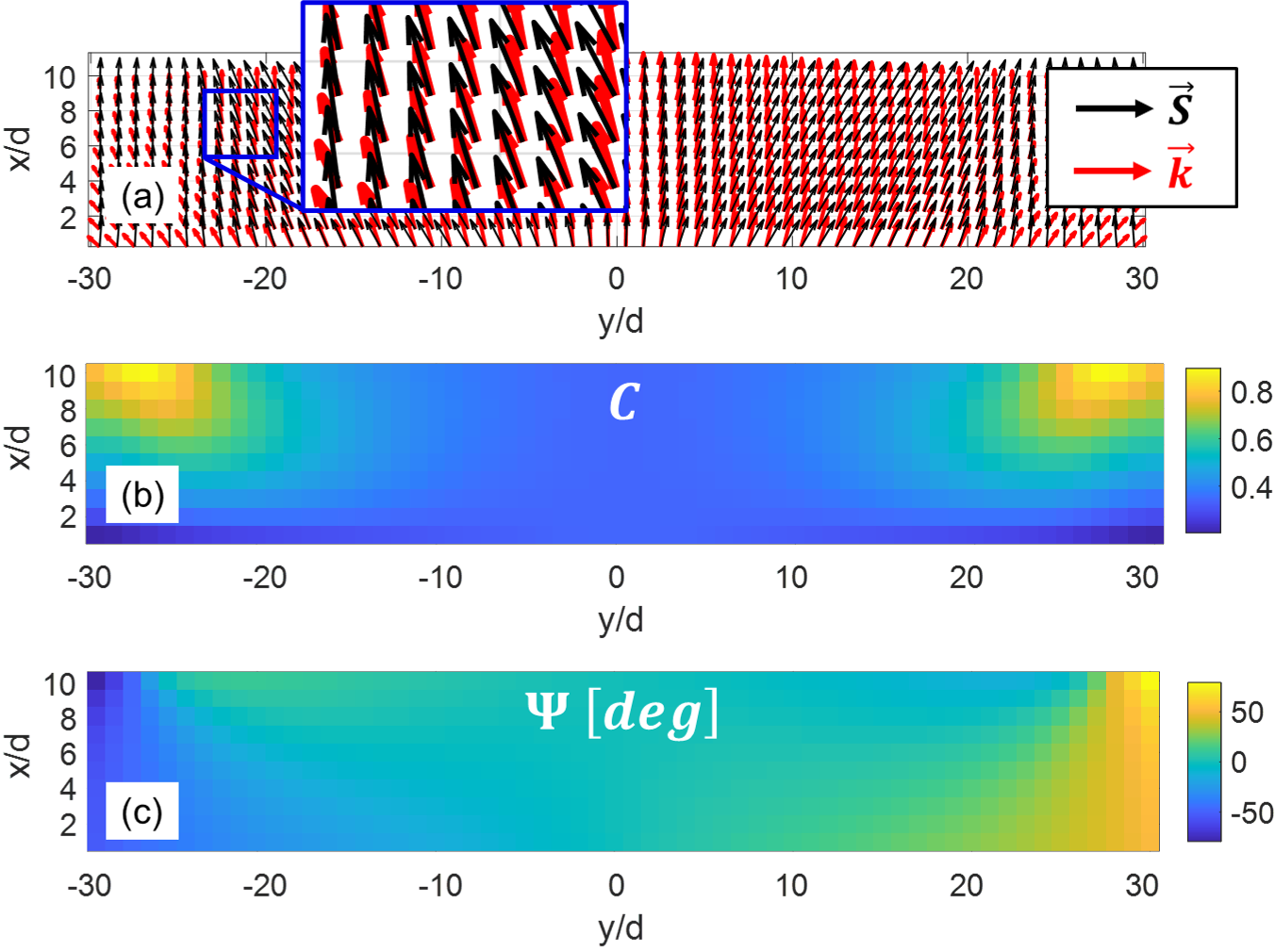}
	\caption{Design parameters for the perfectly matched beam-collimator of Section~\ref{sec:examples_beamCollim}. Spatial distribution of (a) Poynting vector $\boldsymbol{S}$ (black arrows) and wavevector $\boldsymbol{k}$ (red arrows), (b) stretching factor $C$ and (c) rotation angle $\Psi$.}
	\label{fig:example_beamCollim_S_k_arrows_C_Psi}
\end{figure}

\begin{figure*}[t]
	\centering
	\includegraphics[width=1\textwidth]{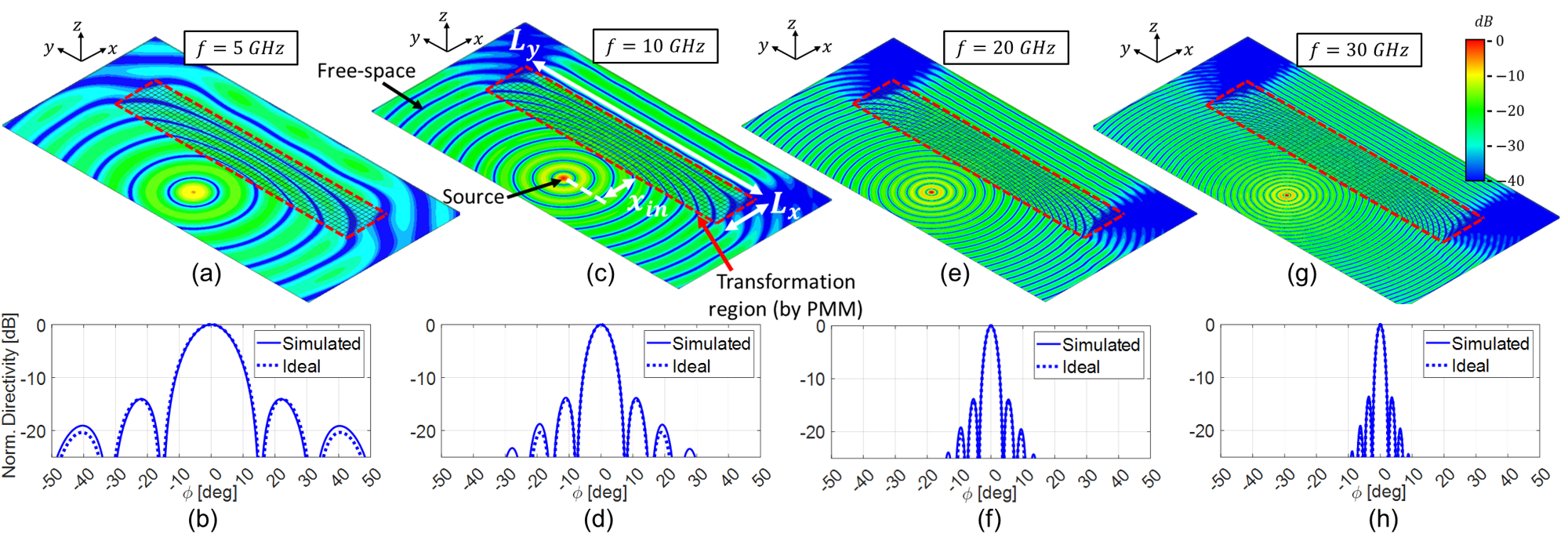}
	\caption{Simulated results for the perfectly matched beam-collimator of Section~\ref{sec:examples_beamCollim} over the 5-30 GHz band. Time snapshot of $E_z$ and simulated and desired radiation patterns at (a-b) 5 GHz, (c-d) 10 GHz, (e-f) 20 GHz and (g-h) 30 GHz. The radiation patterns are computed from the output field of the transformation region. Details for the dimensions of the design example are provided in (c).}
	\label{fig:beamCollim_Ez_radPatt_5GHz-30GHz}
\end{figure*}

\subsection{Perfectly matched beam-collimator \label{sec:examples_beamCollim}}
In this example, we show a broadband beam-collimator designed at $10$ GHz. The device's performance is conceptually sketched in Fig.~\ref{fig:example_PMM_designConcept}. The collimator transforms the field from a line source into a normally-directed plane wave with a trapezoidal amplitude taper. Its dimensions are $d=\lambda/7.2$ (with $\lambda$ calculated at 10 GHz), $x_{in}=10d$, $L_x=10d$ and $L_y=60d$. These dimensions are detailed in Fig.~\ref{fig:example_PMM_designConcept}(a). The transformation region is embedded within free space (see Fig.~\ref{fig:aniMedia_metamaterialPowerPhaseConcept}). Therefore, $\Delta=n_{yy}^2=1$.

For a point source, the input vector power density $\boldsymbol{S_{in}}$ and wavenumber $\boldsymbol{k_{in}}$ at any point $(x,y)$ can be analytically calculated \cite{Harrington:2001}. This calculation, as well as the procedure to determine the local power flow direction and phase, is thoroughly described in the Supplemental Material \cite{Ruiz:2024_PMM_SupplMaterial}. The input and desired output profiles for power density and phase at $10$ GHz are shown in Fig.~\ref{fig:example_beamCollim_powerPhase}. The amplitude taper is chosen such that the power density is maximum along $70$\% of the output interface. Over the remaining $30$\% of the output interface, the power density is linearly tapered to zero. As shown in Fig.~\ref{fig:example_PMM_designConcept}(a), the power at the input and output interfaces of the transformation region is discretized into several points. The points are then connected by straight lines, which provide the power flow direction within each unit cell. A total number of $N=7200$ points are used to discretize the power. As for the phase, we set $\kyNormOut=0$ at the output interface to obtain a normally-directed plane wave. The progression from $\kyNormIn$ at the input interface to $\kyNormOut$ is linear. We fix $\kxNorm^-=0.75$ on the lowest row of the transformation region, at $y=-L_y/2$ (see Supplemental Material \cite{Ruiz:2024_PMM_SupplMaterial}). The resulting local power flow directions and local wavevectors are plotted in Fig.~\ref{fig:example_beamCollim_S_k_arrows_C_Psi}(a). By inserting the obtained power flow directions and wavenumbers into \eqref{eq:pmm_eps} and \eqref{eq:pmm_mu}, we obtain the material parameters within the transformation region. The corresponding stretching factor $C$ and rotation angle $\Psi$, computed through the diagonalization of \eqref{eq:pmm_mu}, are presented in Fig.~\ref{fig:example_beamCollim_S_k_arrows_C_Psi}(b-c).

\begin{figure*}[t]
	\centering
	\includegraphics[width=1\textwidth]{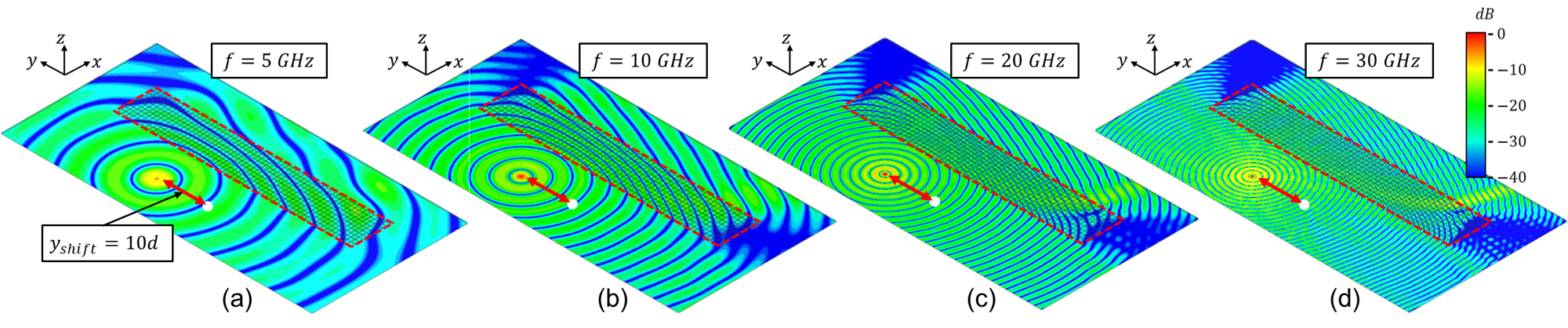}
	\caption{Time snapshot of $E_z$ for a lateral displacement of the source $y_{shift}=10d$ at (a) 5 GHz, (b) 10 GHz, (c) 20 GHz, and (d) 30 GHz.}
	\label{fig:beamCollim_yShift_fields_yMax_allFreqs}
\end{figure*}

\begin{figure*}[t]
	\centering
	\includegraphics[width=1\textwidth]{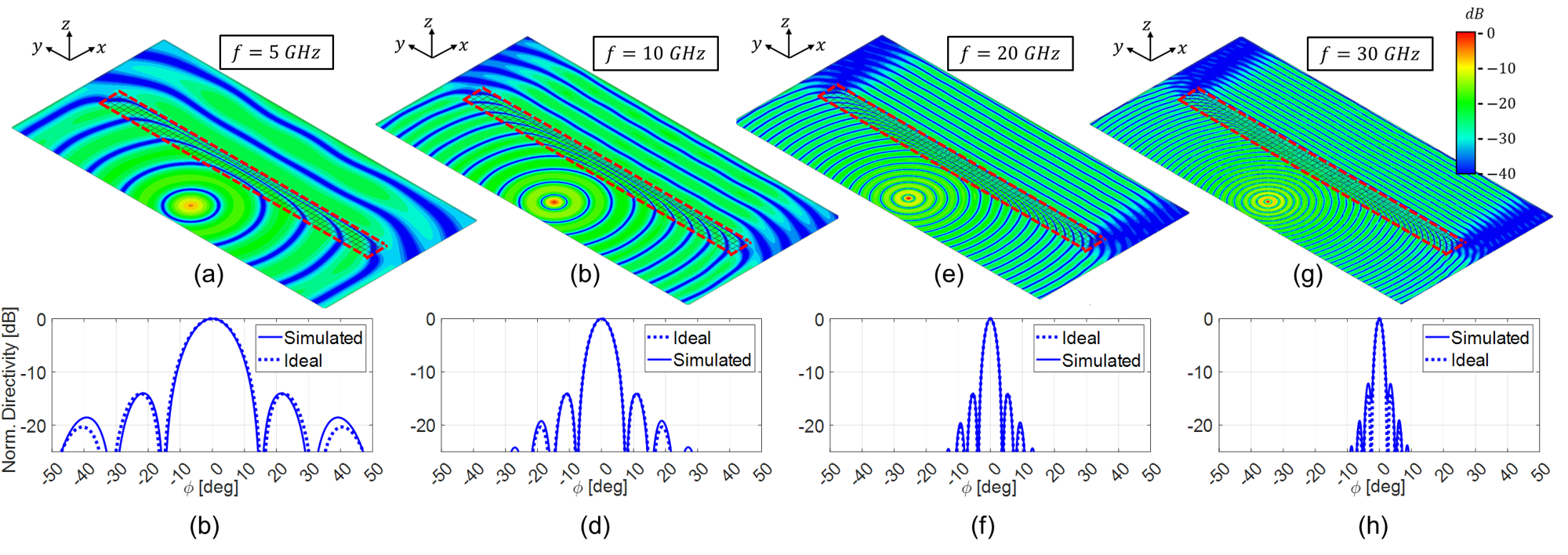}
	\caption{Simulated results for the ultra-thin perfectly matched beam-collimator of Section~\ref{sec:examples_beamCollim_narrow} over the 5-30 GHz band. Time snapshot of $E_z$ and simulated and desired radiation patterns at (a-b) 5 GHz, (c-d) 10 GHz, (e-f) 20 GHz and (g-h) 30 GHz. The radiation patterns are computed from the output field of the transformation region.}
	\label{fig:beamCollim_Ez_5GHz-30GHz_narrow}
\end{figure*}

The beam-collimator is simulated using a commercial full-wave solver \cite{HFSS}. All unit cells in the simulated design are homogeneous, anisotropic blocks whose material parameters yield the stretching and rotation values shown in Fig.~\ref{fig:example_beamCollim_S_k_arrows_C_Psi}(b-c). The resulting performance in the 5-30 GHz band is reported in Fig.~\ref{fig:beamCollim_Ez_radPatt_5GHz-30GHz}. Fig.~\ref{fig:beamCollim_Ez_radPatt_5GHz-30GHz}(a),(c),(e),(g) show a time snapshot of $E_z$. The corresponding desired and simulated radiation patterns in the azimuthal plane, obtained from the field at the output of the transformation region (see Fig.~\ref{fig:example_PMM_designConcept}), are  plotted in Fig.~\ref{fig:beamCollim_Ez_radPatt_5GHz-30GHz}(b),(d),(f),(h). Excellent agreement is observed for the radiation patterns at all frequencies. For a more accurate analysis, the simulated power density and phase profiles at the input and output interfaces at 10 GHz are shown in Fig.~\ref{fig:example_beamCollim_powerPhase}. Agreement with the ideal profiles is clear. Similar near-field plots for the remaining frequencies are included in the Supplemental Material \cite{Ruiz:2024_PMM_SupplMaterial}. As previously mentioned, reflectionless field transformations and non-frequency-dispersive material parameters enable broadband capabilities. In Fig.~\ref{fig:beamCollim_Ez_radPatt_5GHz-30GHz}, one can observe how the line source excitation is collimated without reflection. It is also evident that the field transformation is maintained over a broad band of frequencies. This is quantitatively confirmed by power efficiency, defined here as the power transmitted from the input to the output boundaries of the transformation region. In this case, an efficiency above $97$\% is achieved over the 5-30 GHz band.

PMMs preserve their reflectionless performance under all excitations. However, since we prescribed a field transformation for a particular source, changing the source will result in a different output field. This can be simply demonstrated by displacing the line source of the designed beam-collimator. Fig.~\ref{fig:beamCollim_yShift_fields_yMax_allFreqs} shows a time snapshot of $E_z$ for different frequencies when laterally displacing the source a distance $y_{shift}=10d$ from its original position at $y=0$. One can observe that the structure does not suffer from reflections when shifting the source. In addition, a lens-like beam scanning performance is exhibited by the design. More field plots, as well as normalized radiation patterns, for different values of $y_{shift}$ are provided in \cite{Ruiz:2024_PMM_SupplMaterial}. For a lateral displacement of the source $y_{shift}=10d$, the efficiency remains above $92$\% over the 5-30 GHz band.

\subsection{Ultra-thin beam-collimator \label{sec:examples_beamCollim_narrow}}
In this section, we show reflectionless field transformations involving sharp changes in power flow direction and phase progression. Specifically, we design an extremely thin beam-collimator. The dimensions are those provided in Section~\ref{sec:examples_beamCollim}, except for $L_x$ (see Fig.~\ref{fig:example_PMM_designConcept}(a)). In this case, we fix $L_x=4d$, so that the transformation region has only a 4-unit-cell depth. A line source is again used as the excitation, and the desired power and phase profiles are those shown in Fig.~\ref{fig:example_beamCollim_powerPhase}. Moreover, an approach identical to that in Section~\ref{sec:examples_beamCollim} is adopted to obtain the local power flow direction and phase progression within the transformation region \cite{Ruiz:2024_PMM_SupplMaterial}. The simulated E-field and radiation patterns for the 5-30 GHz band are plotted in Fig.~\ref{fig:beamCollim_Ez_5GHz-30GHz_narrow}. One can observe that, in spite of the extremely thin transformation region, the broadband performance is preserved. This is additionally corroborated by the structure's efficiency, which remains above $98$\% over the bandwidth. The spatial distribution of $\boldsymbol{S}$, $\boldsymbol{k}$, stretching factor $C$ and rotation angle $\Psi$, and the simulated input-output power density and phase profiles are provided in the Supplemental Material \cite{Ruiz:2024_PMM_SupplMaterial}.

\section{Comparison with Transformation Optics (TO) and Perfectly Matched Layers (PML)}
In this section, we complement the information included in Section~\ref{sec:Intro} about the benefits of PMMs over TO techniques. We also summarize the main similarities and differences between PMMs and PMLs.

\subsection{PMMs vs TO}

A more exhaustive comparison can be made with TO structures to highlight PMMs benefits. Conversely to TO, the material parameters of PMMs always have closed-form expressions given by \eqref{eq:pmm_eps} and \eqref{eq:pmm_mu}. These expressions are direct functions of the required local power flow direction and phase progression. Therefore, no coordinate transformations are needed. The most complex step in PMM design is to obtain the local power density direction and phase progression for a given field transformation. However, as shown in Section~\ref{sec:examples_ideal} and the Supplemental Material \cite{Ruiz:2024_PMM_SupplMaterial}, this involves simple analytic calculations. These calculations are arguably simpler than those involved in TO coordinate transformations or mapping to material parameters. 

TO tools are very powerful, but it is difficult to extrapolate and adapt them to inverse design strategies. For instance, MIMO beamformers can be designed by TO techniques only if they are symmetric \cite{Lv:2024_MIMO_TO,Wu:2013_MIMO_TO}. PMM's formulation is more easily adaptable to inverse-design techniques for multifunctional MIMO devices. Indeed, closed-form expressions for PMMs' material parameters can be easily integrated into inverse-design optimizers \cite{Thakkar:2025_TAP,Thakkar:2024_ACES,Thakkar:2023_IMS}. In addition, PMMs can be represented by surrogate models \cite{Thakkar:2023_IMS,Ruiz:2024,Thakkar:2025_TAP}, which dramatically simplify and accelerate their inverse design. As for realization, various types of unit cells are suitable to implement PMMs. An interesting route is the use of all-metal metamaterials \cite{Ruiz:2024,Ruiz:2024_eucap}. Typically, tensor, magnetic materials are considered to be highly dispersive and hard to realize. In contrast, all-metal tensor unit cells can be homogenized as magnetically anisotropic media and engineered to have low frequency dispersion \cite{Ruiz:2024}. This will ensure the broadband field control provided by PMMs. Moreover, using practical unit cells that mimic magnetically anisotropic media allows for the direct realization of PMMs, without the simplification of material parameters often used in TO realizations. In particular, this represents an important advantage over conformal and quasi-conformal TO.

Further, most practical metamaterials are discrete structures. Transformed spaces arrived at by TO are generally continuous, and then discretized for realization. However, PMMs are by nature discretized. Actually, the formulation and design procedure described in this manuscript apply directly to discrete structures.

\subsection{PMMs vs PMLs}
A PMM may be viewed as the lossless counterpart of a PML. They both provide interfaces that remain reflectionless for any angle of incidence and under all excitations \cite{Berenger:1994,Sacks:1995_PML}. However, PMLs are lossy media which do not control fields. They are intended to enclose a computational domain and completely absorb impinging waves, rendering the external boundaries of the domain reflectionless. Designing PMLs involves engineering lossy media (conductivities and/or complex-valued permittivities and permeabilities) with the sole goal of absorbing incoming waves. In PMMs, real (lossless) permittivities and permeabilities are utilized to manipulate propagating waves and transform EM fields.

A well-known drawback of PMLs is that they are not reflectionless when interfaced to discretized spaces \cite{Chew:1996_PML,Oskooi:2008_PML}. In practical simulations, a homogeneous material must be placed between a discrete computational domain and the PML to avoid reflections. On the contrary, PMMs are discrete media whose unit cells can be perfectly matched to any homogeneous surrounding space. An interesting path to further explore is the use of lossy versions of PMMs to truncate discrete computational domains. The effectiveness of this approach has been already shown for inverse-design routines \cite{Thakkar:2025_TAP,Thakkar:2024_ACES,Thakkar:2023_IMS}.

\section{Conclusion}
We introduce Perfectly Matched Metamaterials (PMMs) and their applicability to realizing reflectionless field transformations. PMMs are simple and powerful, and able to independently control the power flow and phase progression of electromagnetic (EM) fields. They are formed from subwavelength, homogeneous, anisotropic unit cells and embedded within a homogeneous, isotropic surrounding medium. For any prescribed field transformation, the required material parameters in each unit cell of the PMM can be analytically calculated. These material parameters inherently provide unit cells that are perfectly matched to each other and the surrounding medium. In conclusion, PMMs are inhomogeneous structures that rely on purely refractive effects to tailor a wavefront. This allows true-time delay field control and the design of broadband devices.

We derived the material parameter properties of perfectly matched media. Next, we described how to relate local power density direction and phase progression with the material parameters of magnetically, anisotropic metamaterials. By applying the perfect-matching conditions to anisotropic metamaterials' unit cells, we derived in \eqref{eq:pmm_eps} and \eqref{eq:pmm_mu} closed-form expressions for the materials parameters of a PMM. Design examples were shown that operate from 5 to 30 GHz, which translates to $\sim140\%$ fractional bandwidth. These structures are formed from ideal, non-dispersive, homogeneous unit cells, and can be approximated using low frequency-dispersive realizations \cite{Ruiz:2024,Ruiz:2024_eucap}.

% Then, we applied the derived formulation to demonstrate broadband capabilities. Design examples were shown that operate from 5 to 30 GHz, which translates to $\sim140\%$ fractional bandwidth. These structures are formed from non-dispersive, homogeneous unit cells, and can be approximated by very low dispersive metamaterials \cite{Ruiz:2024}.

PMMs circumvent some of the most important limitations of Transformation Optics (TO) \cite{Schurig:2006_TO,Kwon:2008,Kwon:2009,Kundtz:2011,Lin:2008_TO,Rahm:2008}. PMMs are devoid of coordinate transformations and their degrees of freedom for field control are maintained throughout/upon their realization by tensor unit cells. This is contrary to approaches such as conformal TO and quasi-conformal TO. These apply optimization and approximation methods to render transformed spaces scalar and non-magnetic. Moreover, PMMs are extremely suitable for inverse-design, unlike TO techniques. We also described the differences and similarities between PMMs and Perfectly Matched Layers (PMLs). While PMLs are used to absorb EM waves, PMMs are used to control them in a lossless manner.
	
The unveiled route to design PMM-based broadband devices performing complex functionalities is still to be explored. Until now, the inverse design of MIMO true-time delay structures using PMMs has been demonstrated \cite{Thakkar:2025_TAP}. In short, PMMs will enable the design of broadband computing metamaterials to perform signal pre-processing and spatial filtering, among other functionalities \cite{Zangeneh:2021}.

\appendix
\section{Perfect matching across $x$-directed boundaries \label{sec:appendix_PMM_xBoundary}}
In the case of all-angle reflectionless transmission across $x$-directed boundaries, a relation is needed between $\overline{\eta}_y$ and $\overline{k}_x$. Following similar steps to those used to derive \eqref{eq:pm_etax_ky_general_modif}, we can obtain,
\begin{equation}
	\left( \frac{1}{\overline{\eta}_y} \right)^2 \left(\frac{\Delta}{\varepsilon_z \mu_{yy}}\right) -
	\frac{\overline{k}_x}{\overline{\eta}_y} \left( \frac{\mu_{xy} - \mu_{yx}}{\varepsilon_z \mu_{yy}} \right) +
	\overline{k}_x^2 \left(\frac{1}{\varepsilon_z \mu_{yy}}\right) = 1.
	\label{eq:pm_etay_kx_general_modif}
\end{equation}
Therefore, the required conditions of the material parameters to enable perfect matching across $x$-directed boundaries are,
\begin{equation}
	\left(\varepsilon_{z} \mu_{yy}\right)_i = n_{xx}^2, \quad
	\Delta_i = \Delta, \quad
	\left(\mu_{xy} - \mu_{yx}\right)_i = \Lambda_{xy}.
	\label{eq:pm_pefectMatching_conditions_Nlayers_xBound}
\end{equation}
For reciprocal media, equation \eqref{eq:pm_pefectMatching_conditions_Nlayers_xBound} reduces to,
\begin{equation}
	\left( \frac{1}{\overline{\eta}_y} \right)^2 \left(\frac{\Delta}{\varepsilon_z \mu_{yy}}\right) + 
	\overline{k}_x^2 \left(\frac{1}{\varepsilon_z \mu_{yy}}\right) = 1,
	\label{eq:pm_etax_ky_reciproc_modif_xBound}
\end{equation}
and the all-angle reflectionless conditions are,
\begin{equation}
	\left(\varepsilon_{z} \mu_{yy}\right)_i = n_{xx}^2. \quad
	\Delta_i = \Delta,
	\label{eq:pm_pefectMatching_conditions_Nlayers_reciproc_xBound}
\end{equation}
A general form of the material parameters is given by,
\begin{equation}
	\boldsymbol{\overline{\overline{\mu}}} = 
	\begin{bmatrix}
		\frac{\left(\Delta + \mu_{xy}^2\right)}{\mu_{yy}} &\mu_{xy} \\ \mu_{xy} &\mu_{yy} \\
	\end{bmatrix}, \quad
	\varepsilon_z = \frac{n_{xx}^2}{\mu_{yy}}.
	\label{eq:pm_pefectMatching_generalForm_muEps_xBound}
\end{equation}
Further, defining $A'=1/\mu_{yy}$, we can write,
\begin{equation}
	\boldsymbol{\overline{\overline{\mu}}} = 
	\begin{bmatrix}
		A' \left(\Delta + B^2\right) &B \\ B &1/A' \\
	\end{bmatrix}, \quad
	\varepsilon_z = n_{xx}^2 A'.
	\label{eq:pm_pefectMatching_ABform_muEps_xBound}
\end{equation}
The material parameters given by \eqref{eq:pm_pefectMatching_generalForm_muEps_xBound} and \eqref{eq:pm_pefectMatching_ABform_muEps_xBound}, which ensure perfect matching across $x$-directed boundaries, are essentially a $90^o$-rotated version of those given by \eqref{eq:pm_pefectMatching_generalForm_muEps} and \eqref{eq:pm_pefectMatching_ABform_muEps} for $y$-directed boundaries.

\begin{acknowledgments}
This work was supported by AFOSR Grant FA9550-24-1-0098.
\end{acknowledgments}

J.R.G. and A.G. contributed equally to this work.

\bibliography{IEEEabrv,prx_PMMs_part1_biblio}

\end{document}